\documentclass[sigconf,nonacm]{acmart}
\AtBeginDocument{%
}
\setcopyright{none}
\begin{document}

\title{``I feel like AI is just everywhere'': A longitudinal qualitative study on parents' perceptions, use, and mediation of GenAI}

\author{Jennica Li, Shirley Zhang, Dakota Sullivan, \\
        Bilge Mutlu, Heather Kirkorian, Kassem Fawaz}
\affiliation{%
  \institution{University of Wisconsin-Madison}
  \city{Madison}
  \state{WI}
  \country{USA}
}
\email{\{jennica.li, hzhang664, dsullivan8, kirkorian, kfawaz\}@wisc.edu}
\email{bilge@cs.wisc.edu}
\renewcommand{\shortauthors}{Li et al.}

\begin{abstract}
GenAI has developed rapidly over the last few years, yet few longitudinal studies exploring how individuals' experiences with GenAI have changed over time exist. Even fewer studies have focused on families with children, despite the opportunity to explore how parents have balanced their needs, their children's development, and the fast pace in which technology changes. To better understand how parents' perceptions, use, and mediation of GenAI shifted over two years, we interviewed 10 parents who had participated in a prior study about their families' experiences with GenAI. Parents had shifted from thinking GenAI was novel to feeling like it was unavoidable, characterizing GenAI as a limited tool to believing GenAI development needed to be regulated, and focusing on their personal mediation strategies to wanting systemic plans to curb GenAI’s long-term consequences. We position our findings within the context of parental mediation of technology and increasing collective mitigation of GenAI-related risks. 
\end{abstract}
\maketitle

\section{Introduction}  
Since debuting to the public in 2022, generative AI (GenAI) systems have seen rapid development and proliferation. From their initial introduction to the general population as a simple chat-based platform~\cite{hu2023chatgpt}, GenAI systems now occupy an influential position in the current sociopolitical zeitgeist~\cite{gottfried2026americans}, yielding comparisons to the cultural shifts that occurred during the dot-com boom and the advent of the smartphone~\cite{eloundou2024gpts}. Yet perhaps what delineates GenAI from these major technological events is the rate at which it has progressed, which has led it to be both lauded as a versatile tool that increases productivity and access to knowledge~\cite{noy2023experimental} and criticized for its seemingly unbridled and sometimes unwarranted proliferation into everyday spaces~\cite{faverio2023key, pi2026push}.

Although researchers acknowledge the dynamism of the GenAI landscape, as evidenced by the impressive breadth of cross-sectional studies produced over the last four years documenting and anticipating the technology’s risks and benefits to users and non-users, longitudinal studies that report more extensive long-term trajectories of GenAI’s impact on individuals are few and far between. Of the longitudinal studies on GenAI that do exist, most focus on how users approach and integrate GenAI in professional or educational settings~\cite{kobiella2024if, ammari2025students}, and often at the expense of positioning GenAI as a static entity that remains generally unchanged between the beginning and the end of the study. Furthermore, while some studies have utilized longitudinal study designs to track within-person change over time during meaningful periods of instability and transition (\textit{e.g.}, becoming a first-year college student~\cite{de2018delicate}), there is a notable absence of longitudinal work examining how families with children have navigated GenAI since its public debut. 

This moment presents an opportunity to examine how families negotiate access to, use of, and communication around emergent and evolving technology to align with children’s developing needs and abilities as they grow older. Parents, in particular, are expected to be arbiters of their children's technology use~\cite{wisniewski2017parental}, and must make decisions with their children's short- and long-term needs in mind~\cite{dumaru2025one}. However, research suggests that GenAI continues to be a mystery to many parents~\cite{mcclain2026teens, yu2024exploring, eira2025parents}, and its relatively recent emergence suggests that the extent of its impacts on children’s development has yet to be fully realized~\cite{diel2026scoping}. 

To address these gaps, this paper presents a qualitative study that explored how nine families with children who were using various GenAI systems in 2024 had shifted their perspectives on and day-to-day use of GenAI about two years later. In the interest of capturing how the evolution of GenAI might be associated with their responses, we asked the parents who participated in our follow-up interviews in 2026 to reflect on what had occurred over the past two years that might have led to their current beliefs, usage– or lack thereof– and mediation (\textit{e.g.}, tech-focused rules, conversation, co-use) of GenAI, and discuss their expectations and concerns about the role that GenAI might play in their children’s futures. Through this study, we aimed to answer the following research questions:
\begin{enumerate}
    \item How, if at all, have parents' perceptions and use of GenAI changed over a two-year period?
    \item How, if at all, have parents’ mediation of GenAI changed over time?
\end{enumerate}

This study makes three following contributions:
\begin{itemize}
  \item First, we offer a novel longitudinal study of how families navigate GenAI during a pivotal two-year period. We compare parents' expectations in 2024, when ChatGPT was still new to families and something they chose to try at home, with their experiences in 2026, when GenAI had become more embedded in their children's schools, devices, and media.
  \item Second, we show that families' use and concern intensified while their mediation did not: parents used GenAI more, grew more pessimistic about its direction, and named more risks to their children, yet none had adopted a mediation strategy by 2026 that they were not already using in 2024.
  \item Third, we offer design and policy implications for family GenAI, highlighting the need for protections that do not rest on household mediation alone.
\end{itemize}

\section{Related Work}
Despite its relative newness as a technology, researchers have made major strides in documenting how GenAI has changed, how GenAI tools are used by different populations, and even some early effects of GenAI on outcomes such as creativity, critical thinking, social skills, identity development, and mental health. However, research on potential longer-term effects of GenAI use is only beginning to emerge. Further, many individuals, including both users and non-users, admit feeling overwhelmed and unknowledgeable about potential promises and pitfalls of GenAI for individuals, families, and society.

\subsection{The Evolution of GenAI}
After Vaswani et al. introduced the transformer architecture in 2017~\cite{vaswani2017attention}, by 2020, large language models were available to developers through commercial APIs~\cite{brown2020language}. In early 2021, DALL-E~\cite{ramesh2021zero} demonstrated text-to-image generation to the public. ChatGPT, released as a free, general-purpose conversational interface in 2022, then became the most prevalent consumer application at the time~\cite{hu2023chatgpt}. Since then, tech companies have used GenAI not as a standalone interface, but as an integrated agent in their products, including Meta AI across Instagram~\cite{meta2024meet}, AI overviews in Google search~\cite{reid2024generative}, and Apple Intelligence in the iOS system~\cite{apple2024intelligence}. These activities have broadened the user population. In the United States, the percentage of adults who reported that they had used ChatGPT before rose from 18\% in 2023 to 44\% in 2026~\cite{gottfried2026americans}. A similar trajectory in use rates can be observed in other countries: compared to about 31\% just a year before, 54\% of UK adults reported using AI tools such as ChatGPT or Gemini in 2025~\cite{ofcom2026adults}, while China's generative AI user base had nearly tripled over one year, from 249 million users in 2024 to 602 million users in December of 2025~\cite{cnnic2026statistical, cnnic2025statistical}.

The rise of GenAI has been met with increasing criticism. Prior work has pointed to the environmental cost~\cite{aczel2026environmental}, data center electricity demand~\cite{zewe2025explained}, and labor displacement~\cite{brynjolfsson2026canaries} associated with GenAI adoption. More relevant to our current study, national organizations focused on youth welfare has emphasized that children are especially susceptible to believing disinformation or being swayed by their interactions with GenAI, which places them at risk of being intentionally deceived and manipulated~\cite{unicefgenai, atabey2025child}. Other research has found that adolescents' use of GenAI for emotional support was positively associated with their reported symoptoms of emotional dysregulation or distress~\cite{bonnesen2026affective}; this finding may be particularly alarming when paired with the statisic that 52\% of US teens reported interacting with a GenAI companion at least monthly in 2025 ~\cite{robb2025talk}. 

While the extant research cannot differentiate whether GenAI use is caused by versus results from emotional distress, correlations such as these, in combination with other concerns noted above, have motivated calls for `ethical' AI. At this time, there is no collective agreement around what constitutes 'ethical' AI \cite{jobin2019global, schiff2021ai}, and many large-scale non-profits and health organizations, including the United Nations Educational, Scientific and Cultural Organization and the World Health Organization, have outlined their own respective guidelines for standardizing the use of GenAI to minimize harm to its users~\cite{unesco2022recommendation, world2024ethics}.

\subsection{Family GenAI Use and Mediation of Technology}
While most GenAI platforms have implemented age restrictions, existing research suggests that children as young as age five have used GenAI chatbots, including ChatGPT and Gemini~\cite{mcclain2025parents}. Additionally, a growing number of GenAI-powered applications are advertised as specifically made for young children~\cite{askie2026}. At this time, there are also limited GenAI-specific guardrails to protect children from the impacts of GenAI. In schools, the majority of conversations around GenAI have been mainly concerned about students using the technology to cheat~\cite{lee2024cheating}, though researchers have also noted individual benefits, including personalized tutoring~\cite{bastani2025generative}. Some laws have been enacted to impose child-specific designs on companion chatbots, including disclosure to known minors, break reminders, and measures to keep sexually explicit content away from children and youth~\cite{rieper2026chatbotlaws}. However, these protective measures are triggered only when an agent can detect that the user is a child. Thus, the responsibility for governing children's GenAI use falls largely on parents. Parents are often expected to negotiate their control over their children’s technology use and access in alignment with children's cognitive development and burgeoning autonomies~\cite{dumaru2025one, wisniewski2017parental}. Notably, research suggests that an increasing number of children across all ages are being exposed to GenAI, yet parents continue to struggle with understanding GenAI well enough to evaluate these interactions~\cite{mcclain2026teens, yu2024exploring}. In our work, we sought to better understand the ways parents mediate their children's GenAI use.

\subsection{Qualitative Longitudinal Studies in HCI}
Longitudinal work is fairly common in the HCI community~\cite{kjaerup2021longitudinal}, and many studies on users' perceptions of technology have recommended longitudinal methods to examine whether attitudes remain stable or shift through continued interactions~\cite{karapanos2009user, de2018phased}. Qualitative longitudinal work, in particular, offers insights around participants' subjective experience over an extended period of time~\cite{thomson2003hindsight}. Existing longitudinal work on GenAI is limited, spanning from a few weeks~\cite{kobiella2024if, long2024not} to about a year~\cite{ammari2025students}, and some studies have focused on a single GenAI tool~\cite{skjuve2022longitudinal}. In tracking their experience over the duration of such a study, participants may have been less likely to notice or recall more granular changes in GenAI performance or function, especially where their scope of reflection was narrowed to a particular agent or platform. This positions GenAI as a static entity, leaving the interaction between dynamic users and dynamic technology underexplored. Researchers have examined families' experiences with GenAI largely through cross-sectional studies~\cite{wen2025families, zhang2025exploring, yu2024exploring, xie2026understanding}. Our approach does not undermine the importance of that work, but complements it by treating both the families and the technologies as changing over the two-year interval we observe.

\section{Methods}

This paper reports follow-up research on prior work first reported by Zhang et al.~\cite{zhang2025qualitative}, which explored families' perceptions, use, and mediation of ChatGPT in 2024; we refer to it as the initial study, or T1, throughout. The present study re-contacted the same families two years later (T2) and, using the original interview transcripts, compares each family’s accounts at both timepoints. To conduct this study, we received permission from the original authors to review and reanalyze the data from the initial study.  

\subsection{Participants and Recruitment}
For the initial study, Zhang et al.~\cite{zhang2025qualitative} recruited 12 families via social media advertisements and e-mail listservs to examine families’ perceptions, use, and mediation of ChatGPT and other GenAI platforms between February and May of 2024 (T1). To participate, at least one parent and one child aged 8 to 18 had to be present for the interview. Additionally, both the parent(s) and child(ren) were required to have some experience using ChatGPT, the most used GenAI platform at the time.

We re-contacted the same 12 families to participate in the follow-up study in April of 2026, explaining that we were interested in understanding their family’s experiences with GenAI over the last two years. 10 parents (7 women, 10 men; age range= 39-53 years) from 9 families completed the study between April and May of 2026 (T2). Because the initial study had intended to explore why families had decided to use GenAI in the household, whereas the follow-up study was more focused on how parents had been navigating their own and their children’s GenAI use and access over a two-year period, we did not require nor ask their children to participate. In lieu of the children’s presence, we instead asked parents to discuss their children’s perspectives and use of GenAI to the best of their knowledge.

All families had at least one parent who had obtained a four-year college degree or higher, and most had a reported annual income of at least \$100000 USD. At the time of the follow-up study, families had children aged 2 months to 18 years of age. All participants were U.S.-based and completed the study virtually. See Table 1 for a more detailed description of participants’ demographics.

\subsection{Procedure}  
Interviews for both timepoints were held remotely over Zoom and lasted between 45 and 90 minutes. During the initial interview, families were asked to describe their initial reactions to ChatGPT, examples of their current and past use of ChatGPT, and any conflicts they encountered while sharing one ChatGPT account with each other. Parents were also asked to discuss any concerns they had with their children using GenAI and other technologies (e.g., social media), and what actions they took to mitigate these concerns.

Prior to the T2 follow-up interviews, we familiarized ourselves with the participants’ T1 data by reviewing the transcripts of their initial interviews and the summaries that had been drafted about each family at T1. During the follow-up interview, we asked parents to begin by describing their family’s current use of technology in general, then asked parents about their current impressions and use of GenAI in particular. We then asked parents to discuss what they knew about their children’s perceptions and use of GenAI and other digital devices, tools, or platforms, before asking about how parents were currently mediating this use. Afterward, we asked parents about how they felt about children using GenAI in general, and what types of GenAI mediation strategies would be most effective. We ended the interview with exploratory questions about GenAI’s increasing presence beyond a screen and asked parents to speculate about what GenAI might look like in another two years’ time. Throughout the interview, we occasionally referenced responses participants had provided during their initial interviews as a way to draw comparisons between their past and present perceptions or behaviors (\textit{e.g.}, \textit{“You previously mentioned that you reviewed the prompts your child input into GenAI platforms. Do you still do this?”}). We also sometimes asked participants to elaborate on what might have motivated a change in perspective or behavior. After the interview, participants completed a brief demographic survey and received a \$25 e-gift card. All study activities were approved by our Institutional Review Board.

\subsection{Data Analysis}
Inductive thematic analysis~\cite{braun2006using} was used at both timepoints to identify themes relevant to the research objectives from the interview transcripts. The T1 interviews were transcribed and coded as part of the initial study~\cite{zhang2025qualitative,zhang2025exploring}; we transcribed and coded the T2 interviews following a similar process. To begin, one coder independently coded a couple of transcripts to develop an initial codebook. Afterward, a second coder used this codebook to independently code the same transcript, then met with the first coder to discuss any disagreements or make any updates to the codebook. This continued until all transcripts were coded, with newly generated codes appended and merged into the final codebook throughout the iterative process. Because both coders had completed consensus coding on all transcripts, we did not calculate inter-rater reliability. Once the codebook was complete, the coders met with an additional author to discuss any emergent themes from the data. Because the initial study had set out to address its own unique set of questions, we maintained separate codebooks for each timepoint. To draw comparisons between T1 and T2, then, we developed sequential analytical matrices~\cite{grossoehme2016analyzing} to explore the trajectories of each family’s reported experiences and identify existing or emergent themes relevant at both timepoints. Although participants had been primed to describe change during the follow-up interview, using matrices allowed us to draw objective comparisons between each timepoint rather than relying solely on participants’ recall. Lastly, we reviewed and finalized the relevant themes presented in this paper.

To maintain confidentiality, we refer to participants by codename that starts with a 'P', then a number that specifies which family they belong to (see Table~\ref{tab:demographics} for family IDs). This number may be preceded by an ‘F’ (e.g., F1) instead of a 'P' if we are referring to the whole family, and followed by a lowercase ‘a’ or ‘b’ (e.g., P9a) to refer to specific parents in families where more than one parent participated. If not already stated in the text, we will include a ‘-T1’ or ‘-T2’ to the end of their codename (e.g., P10\-T2) when attributing quotes to participants to specify whether the quote was from their initial or follow-up interview, respectively.

\begin{table*}[t]
\centering
\footnotesize
\caption{Demographics of the nine families with follow-up data.}
\label{tab:demographics}
\Description{Table 1 reports on the demographics of the nine families that had at least one parent participate in the follow-up study. Column 1 lists each Family ID. Columns 2, 3, and 4 describe parents' reported age, gender, and race, respectively. Columns 5 and 6 list the age of the child(ren) in each family and the type of school their children attended, respectively. Columns 7 and 8 report the highest level of education obtained by parents within each family. Column 9, the final column, lists families' annual incomes. All reported information is based on parent report.}
\begin{tabular}{@{}lllllllll@{}}
\toprule
\textbf{Family ID} & \textbf{P Age} & \textbf{P Gender} & \textbf{P Race} & \textbf{C Age(s) at T2} & \textbf{C Schooling} & \textbf{Parent 1 Edu.} & \textbf{Parent 2 Edu.} & \textbf{Annual Income} \\
\midrule
1    & 50    & W   & Black/White & 15     & Public          & Some college & Bachelor's & \$100,000--\$149,999 \\
2    & 40    & W    & White   & 14, 13, 11, 10, 6, 5, 2 mo    & Public          & Bachelor's   & Master's   & \$100,000--\$149,999 \\
3    & ---   & W   & Asian  & 17, 13, 10   & Public & Bachelor's   & Master's   & --- \\
4    & 53    & W   & White   & 18, 16     & Public          & Master's     & Doctoral   & \$150,000--\$199,999 \\
5   & 46    & W     & White   & 10, 8, 5     & Public          & Doctoral     & Master's   & \$150,000--\$199,999 \\
6   & 39    & W     & White   & 11, 8, 5     & Public          & Doctoral     & Bachelor's & \$75,000--\$99,999 \\
7   & 44/44 & W/M & Asian/White & 13, 11  & Public          & Bachelor's   & Master's   & \$150,000--\$199,999 \\
8  & 49    & M     & White   & 17, 13, 12      & Public          & Doctoral     & Doctoral   & \$200,000 or more \\
9  & 44    & M    & Asian  & 14, 11        & Public          & Doctoral     & Doctoral   & \$150,000--\$199,999 \\
\bottomrule
\end{tabular}
\end{table*}

\section{Results}
Through our matrix analyses, we found that while 8 out of 10 parents (7/9 families) reported that they and their families were using GenAI more often at T2, citing its improved accuracy, efficiency, and versatility for most tasks. At the same time, 7 out of 10 also expressed feeling uncertain or pessimistic about the direction the technology was heading. Every parent, regardless of their general feelings about GenAI, felt that it had progressed faster than they had anticipated. All parents identified at least one concern related to how GenAI might affect their children, and half believed that GenAI would have net negative impacts on their children's futures. To this end, none of the parents introduced a mediation strategy at T2 that they hadn't already employed at T1. 

In analyzing the trajectory of families' experiences with GenAI from T1 and T2, we identified three overarching themes that contextualized parents' changes in perception and subsequent behaviors: 1) GenAI had shifted from being a novel tool to an expected part of life; 2) innovations in GenAI had gone beyond improving its functionality as a tool and was introducing new threats to users; and 3) parents wanted to move beyond at-home monitoring toward greater collective action around mediating children's use of GenAI. 

\subsection{“Not by choice”: From novelty to ubiquity}
At T2, parents characterized GenAI as always present and inescapable. During their initial interviews, parents stated that they could see GenAI providing unique educational or creative opportunities for themselves and their families. Two years later, parents said that GenAI had become less of a novelty that one might explore out of curiosity, and instead an expected presence in daily life: \textit{“I'm using it every day, all the time, whether [I realize it]... and sometimes I realize it”} (P6-T2)

\subsubsection{Concerns around increasing use}
Parents’ foremost concerns involved how the normalization of using GenAI might impact their children’s generation, namely their ability to think critically and adapt to a job market that might value GenAI more than human effort. Many parents discussed how simple and common it was to plug in a complex homework question or essay prompt into a GenAI platform and receive a fully completed assignment. For instance, P1-T2 said: \textit{“You know, not having to read, or actually... not having to think... I feel like the majority of [my child]'s schoolwork is [done by] ChatGPT.”} Under certain circumstances, this use case could be helpful, as described by P7-1 at follow-up: \textit{“And then... for [my child's] algebra homework... [GenAI] has a really great way of, explaining and summarizing... it'll break it into steps... it does a really, a really nice job.”} However, parents felt concerned that this type of use would mean that children would become over-reliant on GenAI whenever they were faced with a task that seemed challenging and lose the opportunities to develop specific academic skillsets because they never forced themselves to practice them: 

\begin{quote}
\textit{“You don't want your kids to be lazy. These [prompts] are things that we should be able to do on our own, be able to write a letter, or things like that... if we're just shunning them off to technology completely, that's failing them in a lot of ways.”} (P2-T2) 
\end{quote}

Reflecting on the large-scale layoffs that had been attributed to GenAI in the last two years, most parents also believed that their children would face difficulties trying to compete against GenAI for roles, and duly so if they entered a job market while lacking essential critical thinking skills, as suggested by P4-T2:

\begin{quote}
 \textit{“...it could put them out of jobs, it could make them unable to think critically, or make their entire generation unable to think critically... It could- it will harm their mental health, it will harm their productivity, it will harm their creativity. You know…there don't seem to be a lot of upsides for my kids.”}
 \end{quote}
 
 At the same time, many parents also believed that, to succeed in their future careers, their children would likely have to be well-versed at using GenAI, which meant it would not serve them to completely limit their interactions with the technology: \textit{“I do think [GenAI] technology absolutely has an important role in their lives, because they'll have to use it for the rest of their lives, and they should have literacy surrounding it.”} (P6-T2) Notably, parent concerns around the job market were mostly unique to T2. Only P9 had briefly mentioned the possibility of job loss due to GenAI during their T1 interview, though P4 had mentioned that they could see GenAI becoming an issue if people began to view GenAI as less of an aide and more of a substitute as well. 

For 6 of 10 parents, another consequence of GenAI's increased use and presence was its impact on the environment. This concern was unique to T2, with one participant,  (P5) specifically mentioning that they had only recently begun to think about it: \textit{“I guess I never really thought about that they need, like, the water, the cooling, like... the environmental impact of it never really crossed my mind until recently.”} These participants mentioned feeling wary about the amount of natural resources (e.g., water, energy) needed to adequately power GenAI. P3-T2 extended their concern to the communities most affected by the construction of data centers: \textit{“You know, they're building data centers left, right, and center. There's no concept of responsibility to the community.”} Four of the participants also emphasized that they were constantly reminded about the environmental implications associated with each prompt or question they input into a GenAI agent. To this end, only P3 and P6 said that they limited their interactions with GenAI as often as possible to not contribute to its resource use. On reflecting about why others might continue to use GenAI, P6 said, 

\begin{quote}
\textit{I don't think we [as a society] realize the environmental impacts [GenAI is] having. I think that there's, like, all of these loaded things that go into it, and I don't know if everyone's on the same page with that.}
\end{quote}

\subsubsection{The loss of agency}
Regardless of how they felt about GenAI's intellectual and environmental impacts, every parent believed that it had become so integrated into the digital landscape that it was almost impossible to abstain from using it altogether, either intentionally or inadvertently. From auto-generated summaries provided by search engines to AI-generated content on social media applications, most parents felt that they or their children would encounter GenAI without trying, as long as they were online. While this did not stop a couple of parents (P3 and P6) from making explicit attempts to limit their and their family’s interactions with GenAI, every parent agreed that their children would grow up alongside GenAI. Whereas parents and children alike had discussed why they chose to opt into using GenAI at T1, several parents at T2 pondered about their inability to control whether they participated in GenAI use: \textit{“[People] will use the AI note-taker, and then I get, like, an email summary of it... It's odd, yeah, like, I don't really consent to it. They chose to do it, and, like, I see [the result].”} (P5-T2). P6 mentioned that, even if they were to be offline entirely, it would not mean they would not encounter GenAI: \textit{“I feel like my kids' schools are using AI-generated handouts and flyers all the time...I feel like AI is everywhere, and I'm exposed to people who are using it to create things.”} Furthermore, some parents mentioned that not using GenAI themselves did not mean they would not be impacted by how others utilized it: 

\begin{quote}
\textit{My husband put in[to ChatGPT], like, what my thoughts were [during our conversation]... then [ChatGPT said] something, like, 'Well, I don't know if your wife is gonna like [your response] based on what you told me about her' … or, like, 'What your wife said here is relevant because of blah blah blah or whatever... That was really weird to just be like, oh, what is this bizarro thing that is telling you what I think about something?} (P4-T2)
\end{quote}

\subsection{``Faster than we can control it'': From `could not' to `should not'}
Most parents no longer viewed GenAI as a limited tool, but instead as a tool that needed limits imposed on its development to minimize societal harm. At T1, parents had focused on the limitations of GenAI’s abilities. At T2, most parents expressed concerns about the unprecedented rate of innovation that GenAI had experienced. Although most parents acknowledged that GenAI had become much more capable in two years, many also wondered at what point GenAI companies would recognize \textit{“the long-term consequences of this unfettered development.”} (P3-T2) 

\subsubsection{Discerning GenAI from reality}
Among parents’ chief concerns was their growing inability to identify GenAI content from human-made content. At T1, parents had been quick to point out common markers of AI-generated content, but parents shifted to talking about how difficult it was to determine whether the content they were looking at had been created by GenAI or a real person at T2. More than half of the parents mentioned that their strategies for spotting AI-generated content (e.g., the number of fingers on someone’s hand) were becoming unreliable: 

\begin{quote}
\textit{“Like, in the beginning, you know, people had, like, extra fingers, like, you could tell very much that things were AI, but I feel like it's starting to get to the point where it's really hard to tell if things are actually real... I don't know how people figure it out, but I'm sure I've seen something that was AI-generated and had no idea, so that's a little bit concerning in that it's harder to tell what's actually real and what's what people have made...”} (P2-T2)
\end{quote}

Though some parents considered GenAI’s improved generative abilities to be positive and felt that the technology could provide utility by producing content that they would \textit{“otherwise not be able to do on their own”} (P2-T2), most parents acknowledged that people might use it to create misleading or harmful videos or images, such as deepfakes: \textit{“AI might create, like, an image of something that happened that didn't actually happen...and you can't always believe your eyes now with the AI, because maybe some of the things you're seeing... it's not real.”} (P5-T2) Several parents admitted that they had already been fooled by GenAI, including P3-T2: \textit{“Do you remember over Christmas, there was this, Christmas tree made of dumplings...and I thought it was so cool...and then my son, my 10-year-old, was like, `Mom, that's AI'.”} Instead of striving toward a technology that might deceive even more people, some parents thought it made sense to implement mandatory disclaimers on any AI-generated content. 

\subsubsection{The value of humanity}
Additionally, more than half of parents also worried that GenAI’s ability to mimic human-made products would diminish the value of human creativity, connection, or perspectives. 4 parents thought normalizing the use of GenAI to produce any form of art would negatively affect artists’ livelihoods by taking away potential work opportunities, minimizing the amount of effort required to produce art, or infringing on their existing copyrights. Additionally, two of the parents, 4 and 18, mentioned that GenAI lacked the capacity to imbue a sense of “humanness” into its content. Though neither parent elaborated on what this entailed, P6 had also mentioned that she evaluated content on “humanness” to determine whether they were AI-generated, alluding to humans’ propensity toward imperfections in their creations: \textit{“I'm kind of looking for... humanness and, like, human error, right? Like, just to kind of figure out is this so polished to a degree where it seems inauthentic?”} Relatedly, five parents said that using GenAI to fulfill a desire for social connection could not and should not compare to the experience of connecting with another person, as it was a tool, not a human: \textit{“If I want to talk with... someone... I prefer a real person. I see those [GenAI] tools as a… problem-solving tool, not a kind of... place [to] put off your emotions, [or] try to.”} (P9-T2) Furthermore, tools for facilitating connection already existed, such as social media platforms: \textit{“... why would [my children] want to interact with a generative AI when they can be talking to their friends, or gaming with people online? And yelling at people all night, you know?”} (P3-T2) 

\subsubsection{Beyond human control}
Finally, 4 parents mused about the likelihood that humans would one day lose control over GenAI if it continued to develop at its current speed. Parents made hyperbolic references to science fiction movies and stories where robots would take over the world, but could anticipate a scenario where GenAI might become too smart to govern: \textit{``And, there's also kind of a sci-fi fiction [element]... AI gonna rebel and then destroy human race [sic]. That's another possibility as well, so… that's what is keeping me from [being] fully optimistic.''} (P9-T2) We noted similar, but fewer comments about this concern at T1: \textit{``And then we're like, 'Oh, dear! We better be quiet [about our complaints], because, you know, [what might happen] when the [GenAI] robots take over the world...''} (P3-T1)

\subsection{“A wild, wild west”: Parental Oversight to Communal Responsibility}
Despite their concerns around GenAI's effects on their children, every parent reported that how they mediated their children's GenAI use remained minimal and unchanged from T1 to T2. Most parents did not monitor their children's interactions with GenAI, even if they had access to their children's accounts or devices. Additionally, some children had global time limits that encompassed any recreational technology use (\textit{i.e.}, not for school), including GenAI use, but multiple parents said they explicitly did \textit{not} consider interactions with GenAI to be screentime: \textit{``I feel like, yeah, screen time is more consumer, or being more of a consumer, and AI is more collaborative?'' }(P2-T2) Most parents had been having conversations with their children about GenAI at T1, but a few parents had mentioned that the frequency of their conversations had increased as they were encountering more instances of GenAI on a daily basis. For example, P7-1 and P7-2 said that while they had not fully explained what GenAI entailed to their two children, they did have brief conversations about identifying AI-generated content with their children \textit{"every other day" }(P7-1) because they were consistently encountering it when using other social media platforms: \textit{"When the content [shows up], we do [too]."} (P7-2) In contrast, P7-2 had stated that they were only talking to their children about GenAI if they wanted to \textit{"just showcase what's possible for the kids"} at T1. 

In most cases, the absence of additional measures was because parents saw no reason to do so because their children had not broken any rules, been exposed to nefarious content, or shown any signs of being negatively affected by GenAI: \textit{" I only [check my child's activity] because I feel like I should... So many people do it, but... I don't really have much to worry about. He's pretty much always with us, and he's a good kid."} (P1-T2) However, parents also attributed their lack of action to the fact that their most pressing concerns felt out of their control. Not only did some parents admit they felt limited in their knowledge of GenAI compared to their children, but felt that the most meaningful ways to mitigate the harm that GenAI might have on their children’s futures should be enacted by schools, GenAI organizations, and government entities. 

\subsubsection{GenAI literacy in schools} 
6 out of 10 parents thought it was important for children to learn about how to use GenAI properly at school. There was no clear consensus around how early into a child’s education these courses should start, though a couple of parents thought that even very young children should learn about GenAI, so long as any version of the technology they interacted with was limited and age appropriate: \textit{“I'm just thinking, like, a kindergartener, you know, starting school, like... If they have access to a device, like, how are they gonna know how to use it and what's safe and what's not safe?”} (P1-T2) When asked about whether their children’s schools had any type of GenAI mediation already in place, 6 parents were unsure, 2 parents said that their children had received some form of GenAI literacy education, and 1 parent said that their children’s school district did not allow students to use GenAI.  

\subsubsection{Product updates and organizational responsibility}
Most of the suggested mediation strategies required technology companies to make updates to their GenAI products. These included developing a child-only version of GenAI, requiring companies to ask individuals to opt into or out of using GenAI, and implementing ‘nudges’ to remind children to reduce their use of GenAI. Relatedly, P3 and P4 felt that it was important to hold GenAI companies accountable for evaluating and reducing when their products could cause harm: 

\begin{quote}
\textit{“I would love it if they were, like, beta-testing everything before they... unleashed it on the world... We don't really have this for anything, but if there was a, you know, a guiding document or philosophy or whatever that was just like... 'do no harm' [and] 'keep protection of kids and families at the center' versus, like, profit generation... I would love for all those things to be true. They're not, they won't be, but it would be nice if they were.”} (P4-T2)  
\end{quote}

\subsubsection{Government-level policies} 
5 parents said they wanted to see more government-level policies to protect children against the negative consequences of GenAI. Whether it was at the local, regional, or federal level, the parents felt that it was important for governments to be involved to set a standard of practice for any GenAI that children interacted with. To this end, parents did not seem to know what exactly these laws or policies might entail, but sometimes referenced policies they had seen in other U.S. states or countries: 

\begin{quote}
\textit{“I think what Grok is doing is terrible, and I think that it's up to governments to ban it and to restrict it. So I think what Indonesia did was great, you know, banning it. I think what Australia and New Zealand are doing to, like, really restrict and crack down on social media and youth is brilliant.”} (P3-T2)
\end{quote}

Additionally, most parents did not want a general blanket ban on GenAI use for children, as they felt children needed to have some access to the technology to be prepared for its presence in their future: 

\begin{quote}
\textit{“Social media is not a tool, it's about communication, and GenAI... it's useful... Luckily or not, I think we're gonna live with generative AI... Our kids did [sic] probably grow with generative AI as their assistance. So...we need to develop some way to... to help them... not ban AI totally.” }(P9-T2) 
\end{quote}

\section{Discussion}
The current study found that over the course of two years, most of the 10 parents reported greater personal and family-wide use of GenAI, but had made minimal changes to their mediation strategies despite new or growing concerns. Parents noted that the whirlwind rise of GenAI meant that it was almost always present in their family's day-to-day lives and could envision how more advanced GenAI tools might benefit its users, but worried that its unrestrained proliferation and development would also lead to adverse effects on their children and society at large. While most parents discussed GenAI with their children and claimed to have some level of oversight over their children's activities, most also wanted support and action from their childrens' schools, GenAI companies, and broader legislative bodies. 

\subsection{Whose choice is it?}
Throughout our analyses, we noticed recurring conversations around what it meant to have 'choice', including having the agency to choose when or how to use and mediate GenAI. Notably, most parents felt that they had few options for mitigating their own and their children's exposure to GenAI. Of the mediation strategies they could decide to use, few, if any, felt like enough to prepare their children for the long-term effects of GenAI on their development, society at large, and the environment. Though it was technically within their power to impose limits on their children's access to GenAI, parents said they lacked the knowledge or tools to do so. Instead, parents felt these could be provided or supplemented by other influential systems in their children's lives (\textit{e.g.}, schools, governments).  

While the context in which parents discussed choice revolved around GenAI in the present study, we note that the concept of agency has been integral to other work exploring families' interactions with and mediation of adjacent technologies. Similar to what we found in the present study, parents may introduce technology to their children on the basis that the device or platform will have short- or long-term benefits for children's development and well-being~\cite{ochoa2020parents}. For example, parents allowed their young children to use screen media not only because they believed that the content could be educational, but also to calm them down~\cite{radesky2016use}. However, parents acknowledge the struggles of balancing their personal beliefs and desires with their children's~\cite{mallawaarachchi2022you, hiniker2016not}. This delicate negotiation may be further complicated as certain technologies become more normalized over time, as children who have minimal or no access to them may be excluded by peers~\cite{albacete2025determinants} or struggle to learn how to use them appropriately when necessary~\cite{livingstone2017maximizing, rodriguez2018study}. Thus, parents may acquiesce and allow their children access to some technologies earlier than they hoped to. 

Importantly, it is difficult to discuss choice without acknowledging how different socioeconomic backgrounds or experiences may affect the level or type of agency parents have around their children's technology~\cite{ames2011understanding}. For instance, families with historically disadvantaged backgrounds are less likely to have access to resources, including those on digital literacy, that would provide knowledge around the various mediation options they could employ~\cite{katz2018connecting}. It is also worth noting that, depending on their backgrounds, parents may have different perspectives on what constitutes appropriate technology use, and thus may be more or less likely to seek out mediation strategies that could support their goals~\cite{clark2012parent}. As the majority of our sample was well-educated and well-resourced, we did not draw any correlations between socioeconomic background and parents' perspectives and actions, but noted certain instances in which their other lived experiences, such as being a teacher, informed their understanding of GenAI. 

\subsection{Societal impacts, societal solutions}

While they did talk about certain short-term implications for their children, parents more often expressed worry about how GenAI would affect their children's long-term futures. Many emphasized that the existing GenAI mediation strategies that were available to them were not enough to prepare and protect their children, and hoped to see more collective efforts around GenAI literacy and regulation to minimize GenAI's effects on their children's critical thinking and social wellbeing, the job market, and the environment. 

We noted that whereas parents seemed to identify different child-centered design choices that GenAI companies could implement into their products, they struggled to extrapolate what types of laws or policies they wanted beyond usually not wanting a blanket ban on GenAI use. It is possible that parents did not feel equipped to make suggestions due to their lack of knowledge of what might be most effective when established as an overarching policy or practice, but had identified a need for greater protections that they could not fulfill. To this end, many countries, including China and the United States, have introduced laws or initiatives aimed at establishing standardized practices and evaluations of GenAI~\cite{ cac2023interim}. Within the United States, numerous local, regional, and statewide governments have also responded to the expansion of GenAI by introducing new policies or moratoriums around GenAI development~\cite{ncsl2025ai}. The long-term efficacy of these policies may be difficult to anticipate, though it is likely that there will need to be continual revisions of any GenAI laws or guidelines to ensure that they are responsive to the most current risks of GenAI.

\section{Limitations \& Future Directions}
We note several limitations in this study and, thus, opportunities for future work. First, the sample, though comparable in size to other longitudinal qualitative studies (\textit{e.g.}, ~\cite{cresswell2007athlete,gray2000qualities,babaria2012m}, comprised mostly White and well-educated participants. Thus, we make no attempt to generalize our findings beyond the present sample, and recognize other samples and populations likely have perspectives and behaviors around technology that would differ from the present sample's~\cite{yardi2012income, ames2011understanding}. Further, as with any longitudinal study, there was attrition from the original study, which further homogenized an already homogeneous sample. On this note, we chose to employ qualitative methods for this study to explore the specific trajectories of the parents involved in this study. To investigate the generalizability of these findings, future longitudinal work could take a mixed-methods approach to obtain the depth provided by qualitative methods and the breadth offered by quantitative approaches.

Second, family perceptions and behaviors were based on parent reports during the follow-up study. The decision to only include parents in this follow-up study was intentional, as we wanted to focus on their unique position as individuals in charge of managing both their and their children's exposure to GenAI. We also chose to mostly report on parents' perspectives for this reason, but acknowledge that references to children's perceptions of behaviors may not be entirely accurate. Future research involving the same study sample may consider re-inviting children to participate if the research questions posed are relevant to the broader family system. Future research might also expand to other age groups, given the early onset of GenAI use for some children and the wide-ranging developmental needs and abilities across childhood and adolescence.

Finally, we believe that there is ample opportunity to further explore how parents' perspectives and management of GenAI continue to evolve alongside the technology. Despite the rate of advancement that GenAI has undergone since 2022, it is still relatively novel and we anticipate that it will continue to develop over time. Additionally, following families with children over a longer period of time would allow researchers to investigate the interplay between children's developmental trajectories and their interactions with GenAI. Given the amount of change that occurred between the initial and follow-up interviews, it may be worth pursuing additional conversations with the study sample at several more bi-yearly intervals to better understand the evolution of GenAI's role and impact on parents and their families.

\section{Conclusion}
Over a two-year period, families' GenAI use increased and they considered GenAI to be evolving faster than parents expected. Participants who once saw GenAI as a tool with built-in limits now thought of it as a technology that needed more limits imposed upon it. Despite growing concerns over GenAI's  developmental, social, and environmental impacts, parents' mediation of their children's GenAI use did not change from two years ago. Instead, the parents looked to technology companies, schools, and governments to create far-reaching and long-lasting change by implementing safe design, promoting GenAI literacy, and enacting sensible regulation. Overall, as GenAI became harder to avoid, the onus of responsibility for mediating children's GenAI use had shifted outside of the home and had become a larger societal issue that required greater collective action. 


\bibliographystyle{ACM-Reference-Format}
\bibliography{main}

\end{document}